\documentclass[rnote]{aa}
\usepackage{graphicx}
\usepackage{color}
\usepackage{natbib}
\usepackage{txfonts}
\newcommand{\cmthree}{cm$^{-3}$}

\newcommand{\kms}{km\,s$^{-1}$}       %km/s
\newcommand{\es}{erg s$^{-1}$}                          %energy cgs
\newcommand{\ecs}{erg cm$^{-2}$ s$^{-1}$}

\newcommand{\um}{$\mu$m}                                 %micron
\newcommand{\molh}{H$_{2}$}                              %H_2, H_2O and H II

\newcommand{\about}{$\sim$}                       %approx
\newcommand{\powten}[1]{10$^{#1}$}
\newcommand{\halpha}{H$\alpha$}                   %H I recombination lines
      
\newcommand{\hbeta}{H$\beta$}

\newcommand{\pbeta}{${\rm P}\beta$}
\newcommand{\pbetalam}{${\rm P}\beta\,1.2818\,\mu$m}
\newcommand{\pgamma}{${\rm P}\gamma$}
\newcommand{\pgammalam}{${\rm P}\gamma\,1.0938\,\mu$m}

\newcommand{\heilong}{He\,I\,1.0830\,\um}         %He I recombination lines
\newcommand{\heishort}{He\,I\,0.5876\,\um}
\newcommand{\av}{$A_{\rm V}$}                     %extinction

\begin{document}

   \title{X-ray and He\,I\,1.0830\,$\mu$m emission from protostellar jets
  }

%   \subtitle{}

   \author{     R.\,Liseau\inst{}        
	}

%   \offprints{}

   \institute{ Stockholm Observatory, AlbaNova University Center, SE-106 91 Stockholm, Sweden \\
   	\email{rene@astro.su.se}
    }

\date{Received date: \hspace{5cm}Accepted date:}

% \abstract{}{}{}{}{} 
% 5 {} token are mandatory

\abstract
 {The high energies of protostellar jets, implied by recent observations of X-rays from such flows, came very much as a surprise. Inferred shock velocities are considerably higher than what was previously known, hence putting even larger energy demands on the driving sources of the jets. The statistics of X-ray emitting jets are still poor, yet a few cases exist which seem to imply a correlation between the presence of \heilong\ emission and X-ray radiation in a given source.}
 {This tentative correlation needs confirmation and explanation. If the jet regions of \heilong\ emission are closely associated with those producing X-rays, high resolution infared spectroscopy can be used to observationally study the velocity fields in the hot plasma regions of the jets. This would provide the necessary evidence to test and further develop theoretical models of intermediately fast ($> 500 - 1500$\,\kms) interstellar shock waves.}
 {The HH\,154 jet flow from the embedded protostellar binary L\,1551 IRS\,5 provides a case study, since adequate IR and X-ray spectroscopic data are in existence. The thermal X-ray spectrum is fed into a photoionization code to compute, in particular, the line emission of He\,I and H\,I and to account for the observed unusual line intensity ratios.}
 {The advanced model is capable of accounting for most observables, but shows also major weaknesses. It seems not unlikely that these could, in principle, be overcome by a time dependent hydrodynamical calculation with self-consistent cooling. However, such sophisticated model development is decisively beyond the scope of the present work.} 
 {Continued X-ray observations, coordinated with simultaneous high resolution infrared spectroscopy, are highly desirable.}

              
 \maketitle

%
%________________________________________________________________

\section{Introduction}

The radiation from HH-objects and collimated jet flows is a manifestation of interstellar shock waves, the typical velocities of which were believed to be $V_{\rm s} \ll 500$\,\kms\ \citep[e.g.,][]{hartigan1987,raga1996,reipurth2001}. Unexpectedly, a number of such objects from embedded protostars have recently been detected at X-ray energies \citep[with~reference~to~HH\,1/2,~HH\,154,~Ser\,SMM1,~HH\,80/81,~TKH\,8,~HH\,158, HH\,210,~respectively]{pravdo2001,favata2002,preibisch2003,pravdo2004,tsujimoto2004,gudel2005,grosso2006}. 

The X-ray spectra have typically a thermal spectral distribution, characteristic of temperatures of several million degrees. This was very surprising, since implied shock velocities ($\gg 300$\,\kms) would be very much higher than what had previously been inferred from spectroscopic observations at longer wavelengths. This indicated that highly significant jet material had escaped detection. As a consequence, there is a potential risk that energy requirements on the jet source, being proportional to the square of the jet velocity, had been seriously underestimated. Proposed energy extraction mechanisms from the protostar-disk system might need to be revised in the light of these new findings.

As in active stars, the {\it continuous} X-ray emission may be accompanied by \heilong\ {\it line} radiation \citep[e.g.,][]{zirin1982}. If that is the case, this line would be a valuable tool to probe the velocity fields of the hottest plasma regions in the jet. Also, in the infrared, extinction problems would be considerably reduced compared to shorter wavelengths and, for consistency checks, recombination lines from hydrogen, i.e. \pbetalam\ and \pgammalam, could also be observed simultaneously in the photometric J-band.

In Table\,\ref{XHeI}, observations of \heilong\ and X-rays toward various flows have been compiled. Evidently, many objects lack the \heilong\ line, but reveal a rich \molh\ spectrum, which is consistent with non-dissociative shocks of low velocity, of the order of 10\,\kms. This is true also for HH\,1G and the Ser\,SMM1 jet, toward which both \heilong\ and X-ray observations have been performed, but neither of which was detected. HH\,2H was detected in X-rays but has apparently not been observed spectroscopically in the 1\,\um\ region. On the other hand, the jet flows HH\,158 (DG\,Tau A) and HH\,154 (L\,1551 IRS\,5) were clearly detected in both \heilong\ and at X-ray wavelengths, lending support to the above proposition. 

It is thus the primary objective of this {\it Research Note} to stimulate further X-ray observations for the investigation of the high energy regions in protostellar jets, alongside with coordinated high resolution near infrared spectroscopy to determine Doppler velocities and line shapes. These data can then be compared with predictions from theoretical shock models.

\begin{table}
\begin{flushleft}
 \caption{\label{XHeI}Protostellar jets with X-ray data and/or NIR spectra}
%\resizebox{\hsize}{!}{   
\begin{tabular}{l ccc l} 
  \hline
  \noalign{\smallskip}
Jet flow& \heilong                & X-rays              & Reference & Comment \\    
  \noalign{\smallskip}
  \hline
  \noalign{\smallskip}      

HH 158  & \phantom{1}$\times$     & \phantom{1}$\times$ & 1, 2      & [Fe\,II]\\
HH 154  & \phantom{1}$\times$     & \phantom{1}$\times$ & 3, 4, 5   & [Fe\,II] \\
HH 210  &                         & \phantom{1}$\times$ & 6         &   \\
HH 1G   & $-\times$               & $-\times$           & 7, 8      & \molh\ + [Fe\,II] \\ 
HH 2H   &                         & \phantom{1}$\times$ & 8         &          \\
TKH 8   &                         & \phantom{1}$\times$ & 9         &          \\
HH 43   & $-\times$               &                     & 10        & \molh    \\
HH 24A  & $-\times$               &                     & 11         & \molh\ + [Fe\,II] \\
HH 25   & $-\times$               &                     & 11        & \molh\   \\
HH 26   & $-\times$               &                     & 11        & \molh    \\
HH 72   & $-\times$               &                     & 11        & \molh    \\
Vel IRS17 & $-\times$             &                     & 12        & \molh    \\
HH 320  & $-\times$               &                     & 11        & \molh    \\
HH 321  & $-\times$               &                     & 11        & \molh    \\
HH 80   &                         & \phantom{1}$\times$ & 13        &          \\
HH 81   &                         & \phantom{1}$\times$ & 13        &          \\
Ser SMM1& $-\times$               &  $-\times$          & 14, 15    & \molh    \\
HH 99-B0& \phantom{1}$\times$     &                     & 16        & \molh\ + [Fe\,II]  \\
  \noalign{\smallskip} 
  \hline
%  \noalign{\smallskip}  
  \end{tabular}
%  }
\end{flushleft}
Notes to the table: $\times =$ detected, $-\times$ means observed but not detected. \\
\molh\ indicates the presence of ro-vibrationally excited \molh\ lines; [Fe\,II] the presence of [Fe\,II] lines; \molh\ + [Fe\,II] the presence of both species along the same line of sight/inside spectrograph slit. When possible, HH-nomenclature follows that of \citet{reipurth1999}. \\ 
References: (1) \citet{takami2002}, (2) \citet{gudel2005}, (3) \citet{itoh2000}, (4) \citet{liseau2005}, (5) \citet{favata2002}, (6) \citet{grosso2006}, (7) \citet{nisini2005}, (8) \citet{pravdo2001}, (9) \citet{tsujimoto2004}, (10) \citet{giannini2002}, (11) \citet{giannini2004}, (12) \citet{giannini2005}, (13) \citet{pravdo2004}, (14) Liseau (unpublished), (15) \citet{preibisch2003}, (16) \citet{mccoey2004}.
\end{table}

\begin{figure}[t]
  \resizebox{\hsize}{!}{
  \rotatebox{270}{\includegraphics{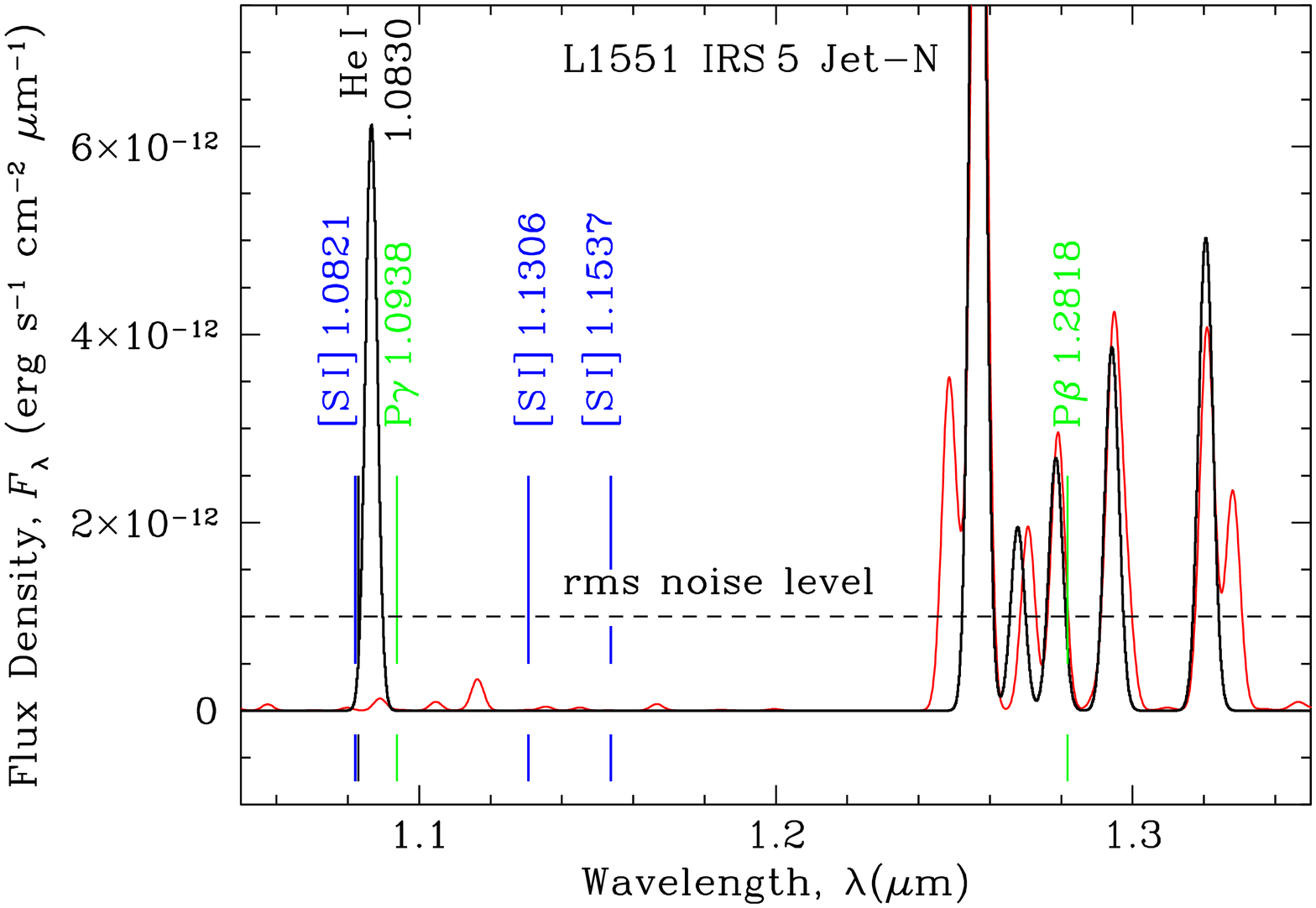}}
                        }
  \caption{The J-band spectrum of \citet{itoh2000} containing the Paschen lines, with the estimated level of the rms noise indicated. The observations are displayed as the black histogram, whereas the smooth red line refers to the fit of the [Fe\,II] spectrum \citep[for~$A_{\rm V} = 5.7$~mag,~][]{liseau2005}. The wavelength positions of the recombination lines \heilong, \pgammalam\ and \pbetalam\ are indicated above and below the spectrum. This applies also to the wavelengths of the transitions to the inverted ground state of [S\,I], i.e. $^1{\rm D}_2 - ^3\!{\rm P}_2$\,1.08212\,\um, $^1{\rm D}_2 - ^3\!{\rm P}_1$\,1.13058\,\um\ and $^1{\rm D}_2 - ^3{\rm P}_0$ at 1.15376\,\um, respectively.
	} 
  \label{IR_lines}
\end{figure}

To date, the best available data are those of the HH 154 flow and we shall first examine the accumulated evidence for this object (Sect.\,2). Some numerical model calculations will be presented and discussed in Sect.\,3. In that section, we will also conclude with some suggestions for future work. 

\section{L\,1551 jet (HH\,154): the data}

\subsection{Infrared spectroscopy}

The examined data are based on the J- and H-band spectroscopy by \citet{itoh2000}. Forbidden lines of ionized iron, [Fe\,II], dominate the observed spectrum and this has been analysed in depth by \citet{liseau2005}. The spectral resolution near $\lambda = 1.0$\,\um\ is quite modest, $\Delta \lambda = 3.6 \times 10^{-3}$\,\um\ ($R = 280$, $\Delta V=10^3$\,\kms), which also corresponds to the accuracy with which wavelengths can be read off Fig.\,5 in the paper by \citet{itoh2000}. 

In Table\,\ref{fluxes}, the line fluxes are given and in Fig.\,\ref{IR_lines}, the jet-spectrum between 1.05\,\um\ and 1.35\,\um\ is displayed. Included is the best fit for the [Fe\,II] transitions of \citet{liseau2005}, shown by the smooth red line, whereas the observations are shown as black histograms (see also Table\,\ref{fluxes}). \citet{liseau2005} tentatively identified the feature near 1.08\,\um\ as He\,I $^3{\rm P}^0_2$ -- $^3{\rm S}_1$\,1.083034\,\um, and in the following, possible other identifications, because of wavelength overlaps within the errors, are briefly discussed.

\subsubsection{[Fe\,II]\,1.0890\,\um}

A little red ``bump'' is visible under the observed line near 1.08\,\um. This bump is due to [Fe\,II]\,${\rm a}\,^4{\rm H} - {\rm b}\,^2{\rm G}$ at 1.08899\,\um, with upper level energy $E_{\rm up}/k > 44\,000$\,K. The matching of the observed line with this transition would require, therefore, a much higher level of excitation than what has been inferred from the other lines in the spectrum, and is therefore inconsistent with the observations. Consequently, [Fe\,II]\,1.0890\,\um\ is disregarded as a viable candidate for the 1.08\,\um\ feature.

\subsubsection{[S\,I]\,1.0821\,\um}

Around 1.08\,\um, atomic/ionic {\it ground state} transitions exist only for neutral sulphur. Relatively close in wavelength to the observed feature is [S\,I]\,$^1{\rm D}_2 - ^3\!\!{\rm P}_2$ at 1.082117\,\um. The other two transitions to the inverted ground state of sulphur are also in the J-band, viz. $^1{\rm D}_2 - ^3\!{\rm P}_1$ at 1.130585\,\um\ and $^1{\rm D}_2 - ^3\!{\rm P}_0$ at 1.1537564\,\um\ (Fig.\,\ref{IR_lines}). None of these latter lines are detected. This is hardly surprising, as sulphur is fully ionized in the jet ($\chi_{\rm S\,I}=10.36$\,eV), which is further supported by the non-detection of the [S\,I] 25.25\,\um\ fine structure $^3{\rm P}_1 - ^3\!{\rm P}_2$ line \citep{white2000}. The observed line near 1.08\,\um\ is therefore not due to neutral sulphur.

\subsubsection{\pgammalam}

Strong Balmer line emission has frequently been observed from the jet \citep[][~and~references~therein]{fridlund2005} and, in view of the high extinction, Paschen lines could be expected to be detectable in the J-band. For instance, using the extinction curve of \citet{rieke1985} for the value of \av\ = 5.7\,mag (derived from the [Fe\,II] spectrum), the attenuation (`reddening') of the \halpha\ flux corresponds to a factor of 70. In contrast, at the wavelength of \pbeta, this extinction amounts to a diminishing factor of only 4. However, based on \halpha\ observations of the jet, the $2.5 \sigma$ feature at 1.2785\,\um\ would appear to be too strong (if real) by a factor of about 3 to be consistent with \pbetalam. As is evident from Fig.\,\ref{IR_lines}, this feature is instead well fit by an [Fe\,II] line.

For a wide range in density, intrinsic Case\,B line ratios are typically \pbeta/\pgamma\,\about\,2 \citep{hummer1987}. Since the observed flux ratio $F(1.2785)/F(1.0866) \ll 1$, and any extinction correction could only decrease this value even further, the observed feature near 1.08\,\um\ is not due to \pgamma. 

\begin{table}
\begin{flushleft}
 \caption{\label{fluxes}Wavelength and flux estimates in the J- and H-band spectra of HH 154 \citep[based~on~observations~by][]{itoh2000} }
%\resizebox{\hsize}{!}{   
\begin{tabular}{crl} 
  \hline
  \noalign{\smallskip}
$\lambda_{\rm obs}^{\,\dagger}$ & $10^{15}\times F_{\rm line}^{\,\ddag}$  & Identification \& $\lambda_{\rm air}$ (\um)\\
(\um)                           & (\ecs)                                  & \citep[FeII:~][]{quinet1996} \\         
  \noalign{\smallskip}
  \hline
  \noalign{\smallskip}      
1.0866	& $26.0 \pm 4.6 $ &  He\,I\,1.083034   \,\,$^3{\rm P}^0\,{\rm 2p} - ^3\!{\rm S}^{}\,{\rm 2s}$ \\
1.2566	& $70.2 \pm 4.6 $ &  [Fe\,II]\,1.256680   \,\,\,${\rm a}^4{\rm D} - {\rm a}^6{\rm D}$\\
1.2785	& $13.2 \pm 5.2 $ &  [Fe\,II]\,1.278776   \,\,\,${\rm a}^4{\rm D} - {\rm a}^6{\rm D}$\\
1.2942	& $19.2 \pm 1.6 $ &  [Fe\,II]\,1.294268   \,\,\,${\rm a}^4{\rm D} - {\rm a}^6{\rm D}$\\
1.3205	& $25.5 \pm 2.1 $ &  [Fe\,II]\,1.320554   \,\,\,${\rm a}^4{\rm D} - {\rm a}^6{\rm D}$\\
1.3272	& $19.9 \pm 4.0 $ &  [Fe\,II]\,1.327776   \,\,\,${\rm a}^4{\rm D} - {\rm a}^6{\rm D}$\\
1.5334	& $20.7 \pm 1.1 $ &  [Fe\,II]\,1.533471   \,\,\,${\rm a}^4{\rm D} - {\rm a}^4{\rm F}$\\
1.5995	& $13.0 \pm 1.7 $ &  [Fe\,II]\,1.599473   \,\,\,${\rm a}^4{\rm D} - {\rm a}^4{\rm F}$\\
1.6435	& $112.3\pm 1.9 $ &  [Fe\,II]\,1.643550   \,\,\,${\rm a}^4{\rm D} - {\rm a}^4{\rm F}$\\
1.6678	& $8.6 	\pm 2.0 $ &  [Fe\,II]\,1.663766   \,\,\,${\rm a}^4{\rm D} - {\rm a}^4{\rm F}$\\
1.6769	& $13.7 \pm 1.7 $ &  [Fe\,II]\,1.676876   \,\,\,${\rm a}^4{\rm D} - {\rm a}^4{\rm F}$\\
1.7045	& $5.0 	\pm 2.0 $ &  $-$ \\
1.7156	& $6.0 	\pm 2.0 $ &  [Fe\,II]\,1.711129   \,\,\,${\rm a}^4{\rm D} - {\rm a}^4{\rm F}$\\
1.7490	& $6.0 	\pm 2.0 $ &  [Fe\,II]\,1.744934   \,\,\,${\rm a}^4{\rm D} - {\rm a}^4{\rm F}$\\
1.8045	& $13.0 \pm 4.0 $ &  [Fe\,II]\,1.800016   \,\,\,${\rm a}^4{\rm D} - {\rm a}^4{\rm F}$\\
1.8134	& $22.0 \pm 6.6 $ &  [Fe\,II]\,1.809395   \,\,\,${\rm a}^4{\rm D} - {\rm a}^4{\rm F}$\\
  \noalign{\smallskip}
  \hline
  \noalign{\smallskip}      
1.0938	& $1.4 	\pm 4.6 $ &  H\,I \pgamma \\
1.2818	& $4.1  \pm 5.2 $ &  H\,I \pbeta \\
  \noalign{\smallskip} 
  \hline
%  \noalign{\smallskip}  
  \end{tabular}
%  }
\end{flushleft} 
Notes to the Table: \\
$^{\dagger}$  Estimated accuracy is $\Delta \lambda = \pm 0.004$\,\um. \\
$^{\ddag}$ Flux calibration from Y.\,Itoh (private communication). 
\end{table}

\subsubsection{He\,I\,1.0830}

For reasons which will become apparent below, we find that \heilong\ is the best identification for the observed 1.08\,\um\ feature. 

Successful models of the infrared [Fe\,II] spectrum of the jet implied \av\,=\,5.7\,mag, where an extinction law of the form $A_{\lambda}/A_{\rm V}= 3.0134-5.621 \lambda + 4.3847 \lambda^2 - 1.5839 \lambda^3 + 0.2181 \lambda^4$, $0.55 \le \lambda \le 2.2$\,\um\, had been adopted \citep{rieke1985}. When the extinction is spectroscopically determined, the uncertainty in \av\ is often dominated not by the measurement errors, but by the uncertainty in the atomic parameters for the transitions from common upper levels (Table\,\ref{fluxes} and Fig.\,1 of Liseau et al., 2005). For instance, the widely exploited line ratio $F(1.64\,\mu{\rm m})/F(1.25\,\mu{\rm m})$ is not better determined than to the range $0.74 - 0.96$ \citep{quinet1996}, where the higher value is based on more recent calculations, and which was the preferred one by \citet{liseau2005}. For the observed ratio, extinction estimates would then fall into the range $A_{\rm V} = 5.7 - 8.6$\,mag (subject to the assumed extinction curve). 

The extinction corrected luminosity of the \heilong\ line is $L_{1.0830\,\mu{\rm m}} = 6 \times 10^{28}$\,\es\ ($D = 140$\,pc and assuming isotropic emission). This corresponds to 20\% of the X-ray luminosity from this region in the jet (see next section).

\subsection{X-rays from the 0.3\,keV $-$ 7.9\,keV band}

\begin{figure}[t]
  \resizebox{\hsize}{!}{
  \rotatebox{00}{\includegraphics{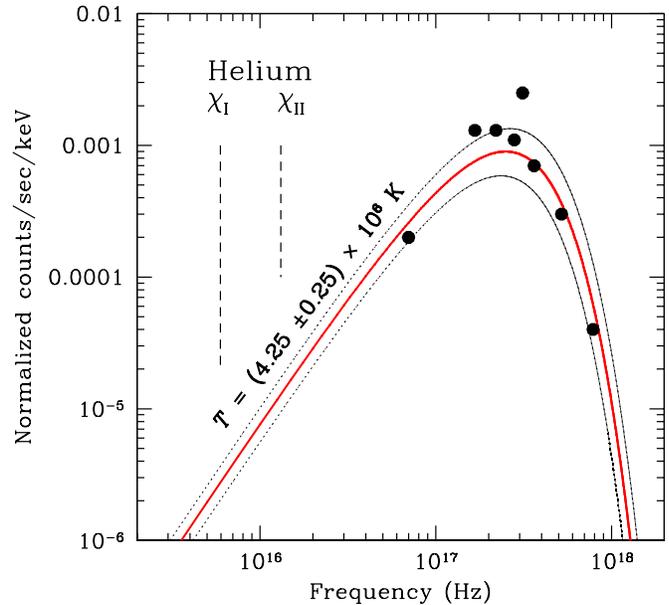}}
                        }
  \caption{The observed X-ray spectrum of the L\,1551\,IRS\,5 jet is shown by the big dots, omitting error bars for clarity. The data are limited to roughly 0.3\,keV $-$ 3.3\,keV. The spectrum appears thermal and is consistent with temperatures of about $4-4.5$\,MK \citep[including~observational~errors,~the~best~fit~MEKAL~model~yields~$4 \pm 2.5$~MK,][]{favata2002}. For the model computations, the spectral shape of the blackbody, shown by the continuous curves, has been used as input spectrum (see the text). The frequencies corresponding to the ionization edges of He\,I and He\,II are shown by the vertical dashed lines.
	}
  \label{xrays}
\end{figure}

At the distance of 140\,pc, the X-ray luminosity is $(2-5) \times 10^{29}$\,\es\ \citep{favata2002,bonito2004}. 
As can be seen in Fig.\,\ref{xrays}, these XMM-Newton data are consistent with a thermal source at temperatures of roughly $4 \times 10^6$\,K. 
\citet{favata2002} give the emission measure of the hot gas as $EM = \int \! (x_{\rm e}\,n_{\rm H})^2 \,{\rm d}V = 1.1 \times 10^{52}$\,\cmthree, yielding an estimate of the source size, the linear scale of which corresponds to a few $\times 10$\,AU for densities of the order of \powten{4}\,\cmthree. 

A large body of observational data exists for the X-ray spot within the (northern) jet \citep{fridlund2005,liseau2005}. The value of the X-ray attenuation, $A_{\rm V}=(7.3 \pm 2.1)$\,mag, is consistent with the visual extinction derived from the infrared [Fe\,II] spectrum. Conforming with this steep extinction gradient is also the observed \halpha/\hbeta-ratio \citep{fridlund2005}. 

The analysis of the X-ray data for protostellar jets (Table\,\ref{XHeI}) has generally been in terms of shock waves and has, as such, frequently exploited analytical formulations for the temperature and pre-shock density, with the shock velocity as a parameter \citep[e.g.,][]{ostriker1988,raga2002}. Common results can be summarized as $T > 10^6$\,K, $V_{\rm s} > 300$\,\kms\ and $n_0 \sim {\rm a~few} \times 10^2$\,\cmthree\ (see the references in Sect.\,1).

\begin{figure}[t]
  \resizebox{\hsize}{!}{
  \rotatebox{00}{\includegraphics{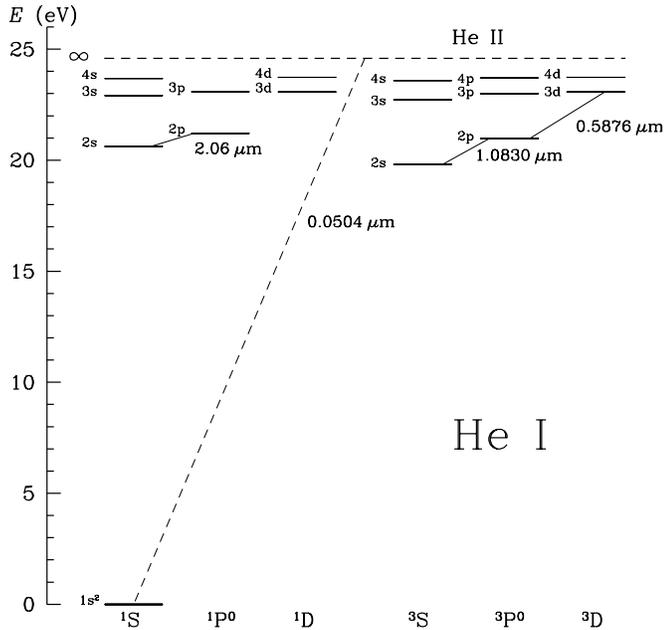}}
                        }
  \caption{Energy level diagramme of He\,I. For clarity, levels for quantum numbers $>4$ have been omitted. The transitions, which are discussed in the text, i.e. 1.0830\,\um\ and 0.5876\,\um\ of the triplet states and 2.06\,\um\ of the singlets, are marked and the wavelength of He\,I ionization is shown next to the slanted dashed line (see also Fig.\,\ref{xrays}).
	}
  \label{HeI_levels}
\end{figure}

\section{Discussion and conclusion}

The overall observational evidence supports the view that the X-ray emission originates from fast shocks in the HH\,154 jet, with velocities in excess of 500\,\kms\ \citep{favata2002,bonito2004,liseau2005,favata2006}. A reasonable scenario features a hypersonic jet, ramming into molecular cloud material which leads to the complete destruction of molecules and grains and the ionization of the gas, which is heated to high temperatures. This hot plasma is radiatively cooled by X-rays, which are capable of ionizing the neutral gas ahead of the shock, altering its state of excitation and potentially also its chemistry. In order to arrive at quantitative estimates of the associated helium and hydrogen recombination line emission, in particular, we have run photoionization models for this scenario. 

The \heilong\ line originates from levels which are more than 20\,eV above the ground (Fig.\,\ref{HeI_levels}). Evidently, the observed X-rays are sufficientally energetic to photoionize helium (see Fig.\,\ref{xrays}) resulting in He\,1.0830 emission upon recombination. But is it possible to explain the fact that the observed line intensity ratio \heilong/\pgammalam\,$>> 1$? Such inverted ratios are generally not revealed by models of jet shocks \citep[e.g.][]{hartigan1987}, but have been observed to develop in supernova ejecta \citep[e.g.][]{meikle1993,pozzo2004}.

For the model calculations we used the program Cloudy \citep[version~96~last~described~by][]{ferland1998}, which determines the temperature structure for a gas of a certain geometry and density, including the transfer of radiation. For the calculations described here, the shape of the X-ray input spectrum is that of a blackbody (Fig.\,\ref{xrays}), scaled by the observed X-ray luminosity, i.e. the inner radius is $r_{\rm in}=\sqrt{L_{\rm X}/4 \pi \sigma T^4}$. The outer radius of the spherical cloud was initially set by the observed X-ray emission measure and the assumed constant density, i.e. $r_{\rm out} \sim (3\,EM/4 \pi)^{1/3} n(\rm H)^{-2/3}$. Solar abundances were assumed for the chemical composition of the gas, hence $n({\rm He})/n({\rm H})=0.1$. 

\begin{figure}[t]
  \resizebox{\hsize}{!}{
  \rotatebox{00}{\includegraphics{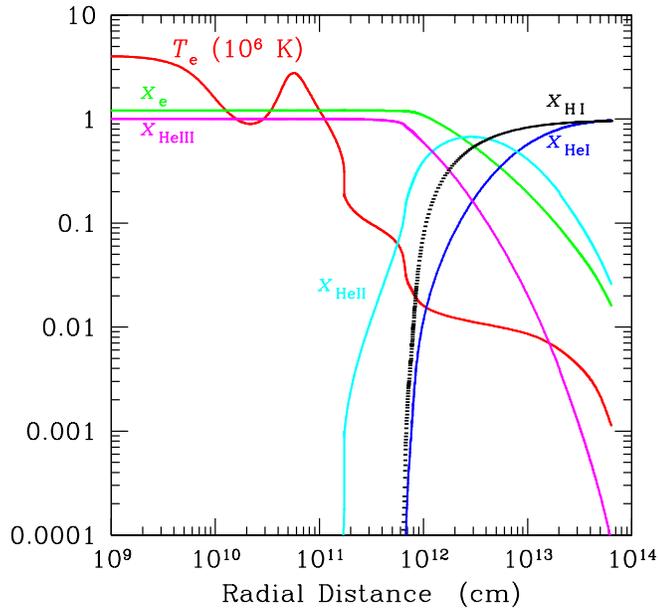}}
                        }
  \caption{Helium ionization structure of gas with constant density $n({\rm H})=10^5$\,\cmthree, being illuminated by a 4\,MK point source. The run of the temperature $T_{\rm e}$ (in units of \powten{6}\,K) is shown in red, that of the electron fraction $x_{\rm e}$ in green. The fractional H\,I density is shown by the black broken line, whereas those of He\,I, He\,II and He\,III in blue, cyan and magenta, respectively. Tests with the previous version Cloudy\,90 gave essentially identical results, except that the temperature structure is smoother. The hump due to `ionization jumps' for a number of species is absent, indicating that this feature does not have any major effect on the results of interest. 
	}
  \label{structure}
\end{figure}

The resulting model for $n({\rm H})= 10^5$\,\cmthree, which is in fair agreement with the observations, is depicted in Fig.\,\ref{structure}. In particular, the ratios \heilong/\pgammalam\ and \heilong/\pbetalam\ are 13 and 7, respectively, i.e. both much larger than unity as has been observed (Table\,\ref{fluxes}). According to the model, with ${\rm e}^-$-collision coefficients from \citet{bray2000}, the excitation of the \heilong\ line is dominated by collisions, so that in this case, the `10830 recombination line' is not really due to He recombination (compare the $x_{\rm He\,II}$ and $x_{\rm He\,I}$ regions in the figure). For purely thermal broadening, the line would be marginally optically thick, i.e. $\tau=1.5$ at the center of the line. The flux of the reddened 2.06\,\um\ line (Fig.\,\ref{HeI_levels}) is predicted not to exceed the 10\% level of that in the 1.0830\,\um\ line.

A hint to the nature of the He\,I excitation is potentially also provided by the \heishort\ line, originating from the state just above that of the \heilong\ line (see Fig.\,\ref{HeI_levels}). The line intensity ratio \heishort/\heilong\ of the model is 0.3, i.e. the predicted (reddened) \heishort\ flux is $4 \times 10^{-16}$\,\ecs. This is in agreement with the observation by \citet{cohen1985}, which resulted in an upper limit, i.e. $F < 6.5 \times 10^{-16}$\,\ecs\ (the units in that paper seem erroneous). Given the information provided by \citet{cohen1985}, it seems not likely, however, that this measurement actually refers to the location of the X-ray source. In addition, the X-ray emission is known to be variable on short time scales \citep{favata2006}. 

The observed \heilong\ luminosity (Sect.\,2.1.4) is a sizable fraction of the X-ray luminosity and it comes hardly as a surprise that the model falls short with regard to $L_{1.0830\,\mu{\rm m}}$. This may not be attributable solely to source variability. In principle, an ad hoc increase of the emitting volume could artificially cure this luminosity problem: for instance, shock models would lead to typically much larger cooling lengths (\powten{15} $-$ \powten{16}\,cm) than what is indicated in Fig.\,\ref{structure} ($< 10^{14}$\,cm). The required increase by an order of magnitude seems consistent with the models of X-ray emitting bow shocks of \citet{raga2002}. Also, as suggested by the referee, a jet with a time variability in the ejection velocity could increase the emitting volume (hence the total luminosity). This volume increase is produced when two successive working surfaces collide with each other.

However, such an ad hoc procedure would lack any physical justification. Therefore, rather than trying `to fix' this poor model, one would like to see a proper time-dependent hydrodynamical calculation for the high velocity jets, accounting for the shocks and the cooling in a self-consistent manner. This is also warranted by the various time scales of the steady state photoionization model (e.g., for thermal equilibrium and hydrogen recombination), which are all very much longer than the dynamical time scale of less than one year. 

On the observational side, one would like to see concerted efforts of X-ray observations and simultaneous intermediate to high resolution ($R \ge {\rm a~few} \times 10^3$) infrared spectroscopy.

\acknowledgements{The thoughtful comments by the anonymous referee are gratefully acknowledged.}

\end{document}